\documentclass[prd,twocolumn,fleqn,showpacs,showkeys,amsmath]{revtex4}

\usepackage{amsfonts}
\usepackage{graphicx}
\usepackage{psfrag}
\usepackage{amssymb}
\usepackage{color}
\usepackage{amsmath}

\newcommand{\rme}{\mathrm{e}}
\newcommand{\rmi}{\mathrm{i}}
\newcommand{\rmd}{\mathrm{d}}

\newcommand{\bfr}{\mathbf{r}}

\renewcommand{\qquad}{\hspace*{25pt}}

\begin{document}

\setcounter{page}{1}%

\title{Axionlike Dark Matter Model Involving Two-Phase Structure\\ and Two-Particle Composites (Dimers)}

\author{A.M.~Gavrilik\footnote{e-mail: omgavr@bitp.kiev.ua}, A.V.~Nazarenko\footnote{e-mail: nazarenko@bitp.kiev.ua}}
\affiliation{Bogolyubov Institute for Theoretical Physics of NAS of Ukraine, \\ 
14b, Metrolohichna Str., Kyiv 03143, Ukraine}%


\begin{abstract}
Within the self-gravitating Bose--Einstein condensate (BEC) model
of dark matter (DM), we argue that the axionlike self-interaction of
ultralight bosons ensures the existence of both rarefied and dense phases
in the DM halo core of (dwarf) galaxies. In fact, this stems from two
independent solutions of the Gross--Pitaevskii equation corresponding to
the same model parameters. The existence of two-phase structure did also appear in
previously studied models with polynomial self-interactions, which actually involve
the truncated expansion series of the axionlike self-interaction.
For a small number of particles, this structure disappears along with
the gravitational interaction, and the Gross--Pitaevskii equation reduces
to the stationary sine-Gordon equation, the one-dimensional antikink solution
of which mimics a single-phase DM radial distribution in the halo core.
Quantum mechanically, this solution corresponds to a zero-energy bound state
of two particles in a closed scattering channel formed by the domain-wall potential
with a finite asymptotics.  To produce a two-particle composite with low positive energy
and a finite lifetime, we appeal to the resonant transition of one asymptotically free particle
of a pair from an open channel (with a model scattering potential) to the closed channel.
Using the Feshbach resonance concept, the problem of two-channel quantum mechanics
is solved in the presence of a small external influence which couples the two channels,
and an analytical solution is obtained in the first approximation.
Analyzing the dependence of scattering data on interaction parameters, we reveal
a long-lived two-particle composite (dimer) possessing a lifetime
of millions of years.  This result is rather surprising and supposes
important implications of dimers' being involved in forming large DM structures.
It is shown that the dimers' appearance is related with the regime of infinite
scattering length due to resonance.
The revealed dependence of the DM scattering length $a$ on the parameters of
interactions can theoretically justify variation of $a$ in the DM dominated galaxies
and its role for large DM structures.
\end{abstract}



\maketitle

\section{Introduction}

Axionlike bosons with low mass and periodic self-interaction belong to
the most popular candidates for the role of dark matter (DM) particles~\cite{Mar16}.
Axions were hypothetically introduced by Peccei--Quinn (PQ)~\cite{PQ77,PQ77b}
as pseudo-Goldstone bosons to resolve the problem of strong charge parity (CP)
in quantum chromodynamics (QCD). Two non-thermal mechanisms for
the axion production in the early Universe were soon proposed, namely
vacuum misalignment~\cite{PWW,AS83,DF83} and cosmic string decay~\cite{Dav86}.
In fact, the thermalization of axions, which were created by the vacuum
realignment in the PQ scenarios with either broken or unbroken symmetry,
seems irrelevant at the early stage because of their initial coherence.
On the other hand, it was argued that the axion component, which appeared during the decay
of topological defects, is thermalized due to self-interactions~\cite{Mar16}.
Besides, gravitational scattering can lead to the re-thermalization of the QCD
axions in the era of radiation dominance, as suggested in~\cite{SY09,SY12}.
This indicates that a system of axions can have huge occupation numbers and
be treated nonrelativistically.
In this regard, the discussion in the literature~\cite{SY09,Dav13,Dav15,GHP15}
on the existence of an axion Bose--Einstein condensate (BEC) led to identifying
any condensate regime with a classical field, regardless of whether the axions
are in the ground state or not. Thereby, it is recognized that the axions in
an occupied state are coherent, so their distribution should be considered from
the point of view of classical field theory or quantum mechanics (in a spirit of
Gross and Pitaevskii)~\cite{Chavanis11}.  Moreover, the wave-based approach
does prevail over the corpuscular one as the particle mass decreases.

It is argued that axions with masses predicted by QCD~\cite{Bor16,KM17} are not
capable to form giant BECs comparable in size to the DM halos~\cite{FMT08,Chavanis2}.
Indeed, the first attempt to describe galactic halos formed by bosons~\cite{Bal83},
either in their ground state (BEC) or in an appropriate isothermal state, resulted
in an extremely small mass of the order of $10^{-24}\ \text{eV}/c^2$. A similar
mass estimate was also obtained in~\cite{Sin} when studying the rotation curves
by considering a giant system of self-gravitating Bose liquid. This means that
there may exist other types of axions with a very small mass, called ultralight
axions~\cite{Mar16}. We also call these (dark) particles {\it axionlike}. Note that
the difference between the masses of QCD axions and ultralight axions can reach
tens of orders of magnitude.

In view of significant mass spread, it is easy to miss the emergence of composites
(complexes of axionlike particles) due to specific nature
of their interaction. In order to fill the gap, we are studying the problem
of axion dimer formation in this paper. 
It is expected that interaction characteristics (e.g., scattering length) differ greatly 
for individual axions and their composites (like dimers).

Noteworthy, a large number of theoretical works are devoted to the study of
the one-sort BEC DM models at vanishing absolute temperature (see, for instance,
\cite{Dav15,GHP15,Chavanis11,FMT08,Chavanis2,Sin,Lee,Hu,Sahni,Ferreira,Bohmer,
Harko2011,axion5}). Their being aimed at the description of astrophysical
phenomena should borrow the chiral cosinelike self-interaction~\cite{PQ77,Wit80}
(or at least its truncated part), and also requires engaging the gravity
(usually treated separately from the unification theory),  the account of which
leads to breaking the inherent symmetry. Note also that there are works, including
experimental ones (see~\cite{Mar16,Brad03,DST06}), that study the interaction of
axions with other substances and their  transformation (say, due to the Primakoff
effect, see~\cite{Prim51,Raff}).

The condensate properties of the gravitating axionlike DM, with both
the leading pairwise contributions to the self-interaction~\cite{Bohmer,Harko2011,axion5}
and the next three-particle corrections~\cite{Chavanis2,GKN20,GN21} being taken
into account, are promising for further exploration and application of axionlike
particles. Having got a number of characteristics, the dilute and
dense phases along with the first-order phase transitions are theoretically
predicted in BEC~\cite{Chavanis2,GKN20,GN21}. Besides, analysis of the effects
and ways to better describe the observables suggest the existence
of moleculelike composites~\cite{GN22} and the relevance
of deformation-based description~\cite{muBose1,muBose2,Nazar}.
Theoretically, these possibilities are considered as very important,
especially when dealing with the dark sector. There are also indications
that quantum entanglement may be involved~\cite{Qent,GN21}.

Thus, there are arguments concerning both the first-order
phase transition at zero temperature with changing the interaction parameters
and its influence on the rotation curves of the DM-dominated galaxies~\cite{GN21}.
The distinct phases could also apparently affect
the state of DM bosonic stars~\cite{Chavanis2}, the merging of which
may produce gravitational waves~\cite{BSmerge1,BSmerge2} as an alternative to black hole
merging~\cite{LIGO}.
Physically, the phase transition in BEC DM is associated with quantum fluctuations
in regions with a relatively high number density of ultralight particles, where
the three-particle effects become significant.

Anyway, the choice of the self-interaction potential is decisive.
Having gained an idea of the nontrivial phase structure of DM with two-
and three-particle interactions and its manifestations in observables~\cite{GKN20,GN21},
we want to show here that the model with the cosinelike interaction
mentioned above should also lead to similar consequences.
Obviously, the already used self-interactions of the polynomial form
now may be treated as the expansion terms of the total potential.
It is important that the cosinelike generalization not only complicates the form
of interaction, but also reduces the number of independent parameters.
We focus on revealing different phases of the DM with axionlike interaction
in the spherically symmetric case, when the main function we find
is the spatial distribution of particles in the BEC.

In general, it is natural to assume that the DM also consists of particles
of different sorts, including composites.
At first glance, composites (``molecules'') would be produced in a dense
environment.
However, as was shown earlier \cite{GN22}, high particle density leads to
particle disintegration triggered by frequent collisions.
But the production of composites may be caused by a large scattering
length of particle interaction.
The very possibility of changing the scattering length is able to shed new
light on the properties and behavior of the initial (elementary) DM particles,
the nature of which have not yet been identified.
What is observed and described, including it in the present study,
is mainly the result of self-interaction (and that of gravitation),
that is, a steady state with a vanishingly small scattering length,
confirmed by numerous models based on the Gross--Pitaevskii
equation~\cite{Harko2011,Chavanis2,GN21,GN22}. Therefore, we need to
explain these distinct particle states separated by an energy gap.

Within qualitative treatment, formation of the simplest ``molecules''
of two and three particles is explored in Ref.~\cite{GN22}.
Here, we turn to scattering processes, using certain analogy
with nuclear processes, and also with experimentally studied phenomena
in atomic BECs in the laboratory~\cite{Grimm10}.
Assuming this to be admissible in the DM as well, we appeal to the quantum mechanical 
formation of a dimer of two particles, borrowing the ideas of the Feshbach resonance and
using two scattering channels~\cite{Fesh62,Joa75,Yam93,Grimm10}.
Although the different channels may be associated with
configurations of internal degrees of freedom of DM particles (a detailed analysis
of which is an independent task), we include auxiliary influences in our consideration
to disclose a plausible mechanism for the formation of bound states during
the Universe evolution.
Thereby, we emphasize the fact that one good potential is not sufficient
to form DM composites in space.

The starting point for studying the dimer formation is the bound state of two
particles held by the axionlike interaction. Then, omitting the gravitational
interaction between a few {\it ultralight} particles, the Gross--Pitaevskii
equation reduces, in fact, to the stationary sine-Gordon equation. 
In order to gain more analytical results, we may restrict ourselves to the 
one-dimensional case that enables reproducing its (anti)kink solution at
zero energy~\cite{Man83}.
In the absence of gravity, the two-phase structure should disappear,
leaving us with one (mixed) phase. Thanks to the analytical solution,
the sine-Gordon potential rewritten as a function of space is,
of course, a domain wall with finite asymptotics. This means that
a pair of particles is in a closed channel with zero energy, and we are
faced with the task of bringing an asymptotically free particle with low
energy into this trap.

As mentioned above, our solution is to admit the existence of
an open channel in which an incident particle is scattered by another interaction
(one of those in which DM particles may participate due to internal properties).
If the particle in this channel has a small positive energy close to
the energy of the bound state (i.e. zero), one can expect a resonant
transition between the open and closed channels in the spatial region of
the trap under a small external influence/force.
This mechanism leads to the appearance of an isolated positive
energy level of a two-particle (compound) system, which becomes
possessing finite lifetime~\cite{Joa75}.
We treat this state as a dimer, whose characteristics are important for
understanding its role in the formation of the DM halo.
According to the Feshbach resonance concept~\cite{Fesh62}, the dimer emergence
at resonance is accompanied by an infinite scattering length, which eventually
depends on the interaction parameters. This fact may also affect
the elucidation of other processes.
Indeed, it was previously assumed that the so-called unitary
regime (of infinite scattering length) could induce instability of the BEC DM halo,
similar to what is observed in the laboratory BEC~\cite{Harko11a,Harko14a}.
But this phenomenon occurs at high density, which we exclude.

The paper is organized as follows. In Sec.~\ref{S2} we show the
existence of two phases of the BEC DM halo core on the base of a pair
of distinct solutions of the stationary Gross--Pitaevskii equation
with axionlike periodic and gravitational interactions at fixed values
of the coupling constants and chemical potential.
The reduction of the Gross--Pitaevskii equation to the stationary
sine-Gordon equation is carried out in Sec.~\ref{Sec3} for the case
of a small number of (ultralight) particles, when the 
gravitational interaction is negligible. 
The exact one-dimensional solution is also discussed there, and its further use outlined.
The Feshbach resonance concept is applied to describe the
axion dimer formation in Sec.~\ref{S4}.
Therein, for convenience of reading, it is divided in five subsections, 
the problem is solved and its various aspects are highlighted.
The necessary preliminaries and the model formulation are given
in Sec.~\ref{S4A} -- it introduces the basic concepts and tools.
The most significant analytical part of study is presented in Sec.~\ref{S4B},
where the quantum mechanics equations of the two-channel problem are solved.
Therein, an isolated energy level of a compound system of two particles is
uncovered and discussed.  Also, analytical expressions for the 
wave functions are given in the first approximation.
The dependence of the scattering length on the interaction parameters
is studied in Sec.~\ref{S4C}. The free parameters of the model are fixed
there, and the Feshbach resonance at zero energy is analyzed.
In Sec.~\ref{S4D}, when considering resonant scattering, the information
about the resonance (and dimer) involving incident particle
with nonzero energy is numerically extracted.
The physical characteristics of the dimer in the context of DM halo
are given and discussed in Sec.~\ref{S4E}.  It is revealed and emphasized that
{\it the lifetime of the dimer is of the order of millions of years}.
The final Sec.~\ref{S5} is devoted to discussion of implications as well as
concluding remarks.

\section{\label{S2}Gross--Pitaevskii Equation for Axionlike DM and Its Solution}

To start with, we formulate stationary macroscopic model of
gravitating Bose--Einstein condensate (BEC) of ultralight bosons
with axionlike interaction $V_{\rm SI}$, restricting ourselves to
the spherical symmetry and the absence of hydrodynamic flows.
Let the BEC with a constant chemical potential $\tilde\mu$ be
described by real function $\psi(r)$ of radial variable $r=|\bfr|$,
with $\psi^2(r)$ defining a local particle density. The behavior
of $\psi(r)$ in the ball $B=\{\bfr\in\mathbb{R}^3|\, |\bfr |\leq R\}$
is determined by the vanishing variation of the energy functional $\Gamma$
with respect to the variation of $\psi(r)$ along with the Poisson equation
for the gravitational potential~$V_{\rm gr}(r)$:
\begin{eqnarray}
\frac{\Gamma}{4\pi}=\int_0^R\left[\frac{\hbar^2}{2m}(\partial_r\psi)^2
-\tilde\mu\psi^2+mV_{\rm gr}\psi^2+V_{\rm SI}\right] r^2\,\rmd r,&&\label{G1}\\
\Delta_r V_{\rm gr}=4\pi Gm\psi^2.&&
\end{eqnarray}

The radial part of Laplace operator $\Delta_r$ and its inverse $\Delta^{-1}_r$
of variable $r$ are
\begin{eqnarray}
&&\Delta_rf(r)=\partial^2_rf(r)+\frac{2}{r}\,\partial_rf(r),\label{Dlt}\\
&&\Delta^{-1}_rf(r)=-\frac{1}{r}\int_0^rf(s)\,s^2\,\rmd s-\int_r^{R}f(s)\,s\,\rmd s;
\label{InvDelta}
\end{eqnarray}
$R$ is the radius of the ball where matter is  concentrated.

The instantonic axionlike self-interaction is chosen here to be
\cite{PQ77,Wit80}
\begin{eqnarray}
V_{\rm SI}&=&\frac{U}{v}\left[1-\cos{\left(\sqrt{v}\psi\right)}\right]
-\frac{U}{2}\psi^2\label{SI1}\\
&=&U \left[-\frac{v}{4!} \psi^4+\frac{v^2}{6!} \psi^6-\ldots\right],\label{SI2}
\end{eqnarray}
where the axion field $\varphi$ is related to the wave function by
$|\varphi|^2=(\hbar^2/m) |\psi|^2$ in the nonrelativistic limit~\cite{Chavanis2}.
Note also that there exists the effective axionic potential~\cite{VV80,Cor16}.

Thus, from the series expansion of $\cos{(\sqrt{v}\psi)}$, we see
that the first term in (\ref{SI2}) corresponds to a two-particle
self-interaction, and the next term $\psi^6$ implies a three-particle
self-interaction.

Note that the works~\cite{Bohmer,Harko2011,Nazar,Zang2018,Kun2020,Craciun2019}
(see also references therein) intensively studied the BEC DM models with
only pair interaction for $v=-|v|$, in the Thomas--Fermi approximation,
when $\hbar\to0$ in Eq.~(\ref{G1}). Switching on the three-particle interaction
enables to observe the two-phase structure of BEC DM that is studied in
\cite{Chavanis2,GKN20,GN21}. For this reason, we expect a similar effect
to occur here.

The axionlike cosine-interaction (\ref{SI1}) is characterized by two constants $U$ and $v$,
which have the dimensions of energy and volume, respectively. In the particle
physics~\cite{VV80,Cor16}, they are related with the axion mass and decay constant
$f_{\mathrm{a}}$ as $U=mc^2$, $v=\hbar^3c/(mf^2_{\mathrm{a}})$. Their
relativistic nature is noted in \cite{Chavanis2} in the context of nonrelativistic
model of axion stars. However, in astrophysical applications, these constants may have
other meaning and values, which will be discussed below.

To analyze the general properties of the model, we reformulate it in dimensionless
variables:
\begin{eqnarray}
\xi=\frac{\sqrt{mU}}{\hbar}\,r,&\quad&
\chi(\xi)=\sqrt{v}\,\psi(r),\label{mp1}\\
\xi_B=\frac{\sqrt{mU}}{\hbar}\,R,&\quad&
A=4\pi\frac{G\hbar^2m}{U^2v},\label{mp2}\\
u=\frac{\tilde\mu}{U},&\quad& \nu=1+2u+2A\Phi_0,
\label{mp3}
\end{eqnarray}
where $\nu$ plays the role of effective chemical potential, which
absorbs the constant term of axion interaction and the
gravitational potential at the origin $\xi=0$, namely
\begin{equation}
\Phi_0=-\int_0^{\xi_B}\chi^2(\xi)\,\xi\,\rmd\xi.
\end{equation}
In our study, $\nu$ is regarded as a free variable parameter,
due to arbitrariness of $u$.

The model equations in terms of the wave-function $\chi(\xi)$ and
auxiliary gravitational potential $\Phi(\xi)$ read
\begin{eqnarray}
&&\left(\Delta_{\xi}+\nu\right)\,\chi-2A\Phi\chi-\sin{\chi}=0,\label{feq1}\\
&&\Phi(\xi)=-\frac{1}{\xi}\int_0^\xi \chi^2(s)\,s^2\,\rmd s+\int_0^{\xi}\chi^2(s)\,s\,\rmd s,
\label{feq2}
\end{eqnarray}
where $\Delta_\xi\Phi(\xi)=\chi^2(\xi)$ is satisfied, and $\Phi(0)=0$.

To obtain a finite and stable solution $\chi(\xi)$, the non-linear
Eqs.~(\ref{feq1})-(\ref{feq2}) should be (numerically) integrated under
the following conditions: $\chi(0)<\infty$, $\chi^\prime(0)=0$,
$\chi^{\prime\prime}(0)<0$. For given $A$ and $\nu$, the finite
initial value $\chi(0)=z$ should be positive solution of the transcendental
equation
\begin{equation}\label{icon}
2Az^2+\left(\nu-\frac{\sin{z}}{z}\right)(\nu-\cos{z})=0
\end{equation}
which is derived by substituting $\chi(\xi)=\chi(0)+\chi^{\prime\prime}(0)\xi^2/2$
into (\ref{feq1})-(\ref{feq2}) for $\xi\to0$ and by finding $\chi^{\prime\prime}(0)$.
Note that the requirement $\chi^{\prime\prime}(0)<0$ is equivalent to
imposing $\nu>\sin{z}/z$ for positive $\nu$.

The absence of a solution $z$ for a given pair $(A,\nu)$ means that
$\chi(\xi)=0$ everywhere. It happens for $\nu>\nu_{\mathrm{max}}$,
where $\nu_{\mathrm{max}}(A)$ is also found numerically from
(\ref{icon}). For $\nu<\nu_{\mathrm{max}}$, two branches of
$\chi_0(\nu)$ can occur, which indicate the existence of
two regimes  and a first-order phase transition in the model.

\begin{figure}[htbp]
\includegraphics[width=6.2cm,angle=0]{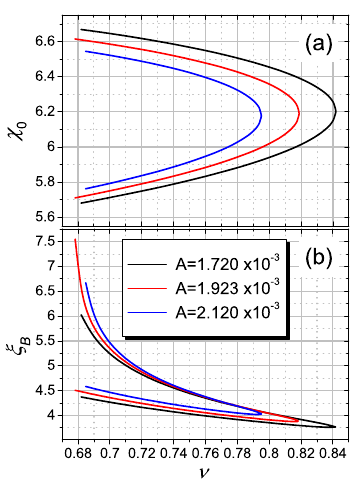} 
\caption{\label{chi0}The initial value $\chi_0=\chi(0)$ (a) and the first
zero $\xi_B$ (b) of $\chi(\xi)$ versus parameter $\nu$ for different $A$.
The curves are limited from the right by the values
$\nu^{\mathrm{black}}_{\mathrm{max}}=0.841954176$,
$\nu^{\mathrm{red}}_{\mathrm{max}}=0.818967072$,
$\nu^{\mathrm{blue}}_{\mathrm{max}}=0.795228679$.
Each rightmost point $\chi_s=\chi_0(\nu_{\mathrm{max}})$
divides the corresponding curve into upper and lower branches
identified with different phases.
At $\nu>\nu_{\mathrm{max}}$ one has $\chi_0=0$ and $\chi(\xi)=0$.}
\end{figure}

Let us emphasize that the magnitude of parameter $A$ plays a crucial
role for subsequent implications. Assuming that gravity is weaker than
the axion self-interaction, we take the parameter $A\gtrsim10^{-3}$. Then,
Eq.~(\ref{icon}) leads to two independent solutions for $z>\pi$. Indeed,
choosing $A$ as in Fig.~\ref{chi0}, the upper branch of $\chi_0(\nu)$
corresponds to $\chi_0\in[\chi_s;5\pi/2]$, while the lower branch of
$\chi_0(\nu)$ gives us values of $\chi_0$ within the interval $[3\pi/2;\chi_s]$,
where the separating value $\chi_s=\chi_0(\nu_{\mathrm{max}})\lesssim2\pi$.

Therefore, there exist two independent solutions $\chi^{(\alpha)}(\xi)$,
$\alpha=1,2$, of the set of Eqs.~(\ref{feq1})-(\ref{icon}) for
the same parameters $A$ and $\nu$.
They are characterized by $\chi^{(\alpha)}_0$ and $\xi^{(\alpha)}_B$,
which belong to different branches ($\alpha=1,2$) as in Fig.~\ref{chi0}.
This means that any space average $F(\nu)$ in the statistical description,
for instance, {\it mean particle density}
\begin{equation}\label{siggen}
\sigma(\nu)=\frac{3}{\xi^3_B}\int_0^{\xi_B}
\chi^2(\xi) \xi^2\,\rmd\xi,
\end{equation}
for fixed $A$ and variable $\nu\leq\nu_{\rm max}$, also consists of two branches
$F^{(1)}(\nu)$ and $F^{(2)}(\nu)$, e.g. 
\begin{equation}
\sigma^{(\alpha)}(\nu)=3 \left(\xi^{(\alpha)}_B\right)^{-3}\int_0^{\xi^{(\alpha)}_B}
\left[\chi^{(\alpha)}(\xi)\right]^2\,\xi^2\,\rmd\xi.
\end{equation}
Then the graph of the function $F(\nu)$ is the union of the graphs
for $F^{(1)}(\nu)$ and $F^{(2)}(\nu)$ so that
$F^{(1)}(\nu_{\rm max})=F^{(2)}(\nu_{\rm max})$ by construction.
But it is convenient for us to continue the use of the definition (\ref{siggen}),
keeping in mind the need to take into account different branches.

Contrary to the expectation of a weak axion field ($\chi_0<\pi$) near the true
vacuum~\cite{Wit80}, the situation looks different in the nonrelativistic
model of DM with Newtonian interaction. Also, there is no invariance there
under the global transformation $\chi\to\chi+2\pi$ (it is violated by gravitation).
Besides, such a discrepancy is related with the consideration of the condensate
in a finite volume (of galactic DM halo). We might expect some distinctions when
describing gravity as a space-time geometry.

Indeed, Fig.~\ref{chi0}(b) demonstrates the value of (first) zero $\xi_B$ of
oscillating function $\chi(\xi)$ (that is, $\chi(\xi_B)=0$), which limits
the system size in our model and is found by integrating
Eqs.~(\ref{feq1})-(\ref{feq2}) for given $A$ and $\nu$.

\begin{figure}[htbp]
\includegraphics[width=5.8cm,angle=0]{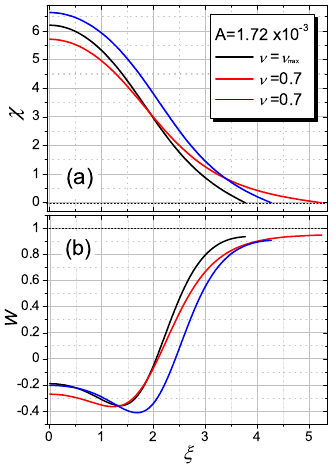} 

\vspace*{-1mm}
\caption{\label{chi}A particular form of the wave function $\chi(\xi)$ (a) and
the effective potential $W(\xi)$ (b) for the fixed parameters $A$ and $\nu$.
Blue and red lines are obtained with the same $A$ and $\nu$, but differ in
the initial values $\chi_0$, see Fig.~\ref{chi0}(a). The value of $\chi_0$
for blue curves belongs to the upper branch of $\chi_0(\nu)$ in
Fig.~\ref{chi0}(a), while the red lines are plotted using $\chi_0$ of
the lower branch of $\chi_0(\nu)$. Black curves correspond to a single state
at $\nu=\nu_{\rm max}$.}
\end{figure}

The typical solutions for the wave function $\chi(\xi)$ are presented in
Fig.~\ref{chi}(a), where $\xi\in[0;\xi_B]$. They describe a core with a finite
magnitude of the particle (and mass) density at the center $\xi=0$. Besides,
an additional information may be extracted through introducing the effective
potential $W$ [see Fig.~\ref{chi}(b)] as a function of radial variable $\xi$,
namely
\begin{eqnarray}
&&W(\xi)=2A\Phi_0+2A\Phi(\xi)
+\frac{\sin{\chi(\xi)}}{\chi(\xi)},\label{sch1a}\\
&&\left[-\Delta_\xi+W(\xi)\right]\chi(\xi)=\varepsilon\chi(\xi),\label{sch1b}
\end{eqnarray}
when Eq.~(\ref{feq1}) is re-written in the form of Schr\"odinger equation
with the ``energy'' $\varepsilon=1+2u$, see Eq.~(\ref{mp3}).

Fig.~\ref{chi}(b) shows the forms of potential $W(\xi)$. The black curve
describes the effective potential of the mixed (single-phase) state at
$\nu=\nu_{\rm max}$. The blue and red curves correspond to the effective
potentials of two phases for the same parameters $A$ and $\nu<\nu_{\rm max}$,
but obtained from the solutions $\chi(\xi)$ with different $\chi_0(\nu)$
and $\xi_B(\nu)$ belonging to different branches in Fig.~\ref{chi0}.
The different positions and depths of the minima of these potentials confirm
the existence of two macroscopic states in the axion system. The particles being
in one of the two phases is conditioned by the applied
factors e.g. pressure~\cite{GKN20,GN21}.

Outside the system at $\xi>\xi_B$, where the matter is absent, $W(\xi)$
is continuously extended by the gravitational potential of the form
$1-2A{\cal N}/\xi$, where ${\cal N}$ is the total number of particles
in the ball $\xi\leq\xi_B$.

Let us estimate the characteristics of our model, which are related
with the dimensionless quantities (\ref{mp1})-(\ref{mp3}).  
For this aim, we refer to the model from \cite{GKN20}, which dictates
to separate the approaches to describing the core and the tail of the DM halo
due to different role of self-interaction in relatively dense and
rarefied regions.
It is worth noting that recent astronomical measurements
indicate spatial fluctuations in the density of cold DM (around the quasar)
on the scale of 10 kpc~\cite{ALMA}, which may be associated with
an oscillating tail of the condensate wave function, rather than
the mentioned smooth tail.

Focusing on the phenomena in the core, we need to
reproduce the size scale $r_0$ and the central mass density $\rho_0$,
which define $r=r_0\xi$ and $\rho(\xi)=\rho_0\chi^2(\xi)/\chi^2_0$.
At the same time, this is needed to control the parameters $U$ and
$v$ in (\ref{SI1}).

As it was stated in Ref.~\cite{Chavanis2}, models for describing
compact objects (such as axion stars) and cosmological models lead
to different parametrizations of axions. First of all, this concerns
the different mass ranges.
While cosmological models constrain the axion mass as
$10^{-7}\,\text{eV}<mc^2<10^{-2}\,\text{eV}$, another kind of
models suggests that $mc^2\sim10^{-22}\ \text{eV}$, commonly
attributed to fuzzy DM~\cite{Hu,Peebles}. Physically, the choice
of a smaller particle mass ensures the formation of certain
structures in the Universe~\cite{FMT08},
some of which we are trying to describe. It motivates us to fix
$m\sim10^{-22}\ \text{eV}\,c^{-2}$. Although the parameter $U$
in the chiral models is set proportional to $mc^2$ \cite{Cor16},
we do not declare such an identity in the nonrelativistic case,
and admit only that the value of $U$ provides the predominance
of axion repulsion over gravitation at relatively large magnitude
of axion field.

Besides, we have to determine $f_{\mathrm{a}}$ in order to specify
$v=\hbar^3c/(mf^2_{\mathrm{a}})$. Although
$10^{18}\,\text{eV}<f_{\mathrm{a}}<10^{21}\,\text{eV}$ in cosmology~\cite{PWW},
the decay constant $f_{\mathrm{a}}$ may be appearing larger
in the models with ultralight particles~\cite{Chavanis2}.

Thus, combining the definition $v=\hbar^3c/(mf^2_{\mathrm{a}})$ and
the relation $m \chi^2_0=v\rho_0$ resulting from~(\ref{mp1}), we obtain that
\begin{eqnarray}
f_{\mathrm{a}}&\simeq&3.304\times10^{19}\,\text{eV}\,
\left[\frac{\rho_0}{10^{-19}\,\text{kg}\,\text{m}^{-3}}\right]^{1/2}
\nonumber\\
&&\times\left[\frac{mc^2}{10^{-22}\,\text{eV}}\right]^{-1}
\left[\frac{\chi_0}{2\pi}\right]^{-1},
\end{eqnarray}
where the central mass density $\rho_0$ is taken to be of the order of
$10^{-19}\,\text{kg}\,\text{m}^{-3}$, while a mean mass density
is assumed to be of the order of $10^{-20}\,\text{kg}\,\text{m}^{-3}$
as usual~\cite{Harko2011,Chavanis2}.

Using the second relation of (\ref{mp2}) and $m \chi^2_0=v\rho_0$, we find:
\begin{eqnarray}
U&\simeq&2.145\times10^{-29}\,\text{eV}\,
\left[\frac{\rho_0}{10^{-19}\,\text{kg}\,\text{m}^{-3}}\right]^{1/2}
\nonumber\\
&&\times\left[\frac{A}{2\times10^{-3}}\right]^{-1/2}
\left[\frac{\chi_0}{2\pi}\right]^{-1}.
\end{eqnarray}
The characteristic scale is defined here as $r_0=\hbar/\sqrt{mU}$ and equals to
\begin{eqnarray}
r_0&\simeq&0.138\,\text{kpc}\,\left[\frac{\rho_0}{10^{-19}\,\text{kg}\,\text{m}^{-3}}\right]^{-1/4}
\left[\frac{mc^2}{10^{-22}\,\text{eV}}\right]^{-1/2}
\nonumber\\
&&\times
\left[\frac{A}{2\times10^{-3}}\right]^{1/4} \left[\frac{\chi_0}{2\pi}\right]^{1/2}.
\label{r00}
\end{eqnarray}
Such $r_0$ is appropriate for estimating the size of the central part
of the DM halo as $R=r_0\xi_B$, but should be fitted together with
the total mass $M$.

To justify the first-order phase transition in the model,
one has to develop a statistical approach, which is omitted
here. Nevertheless, the discontinuous change in particle
density~(\ref{siggen}) at zero temperature is expected to be caused
by a change in the long-wave part of pressure~\cite{GKN20}
\begin{equation}
\Pi=-\frac{3}{\xi^3_B} \int_0^{\xi_B} \left[(\partial_r\chi)^2-\nu\chi^2\right] \xi^2\,\rmd\xi,
\end{equation}
which also consists of two branches, as stated above.

Clearly, the effect of different DM phases on the observables, as well as on
the rotation curves, also deserves a separate study.

\section{\label{Sec3}Sine-Gordon Equation and a Bound State}

Let us analyze the ground state of the DM halo core by turning to
a one-dimensional model with the coordinate $\xi\in[0;+\infty)$,
when gravity is absent and only the axion self-interaction plays a key role.
This means that Eq.~(\ref{sch1b}) under the 
simplifications
\begin{equation}
A=0,\qquad \varepsilon=0, \qquad \Delta_{\xi}\to\frac{\rmd^2}{\rmd\xi^2}
\end{equation}
reduces to the stationary sine-Gordon equation:
\begin{equation}\label{SG1}
\left(-\frac{\rmd^2}{\rmd\xi^2}+W\right)\chi=0,\qquad
W=\frac{\sin{\chi}}{\chi},
\end{equation}
which is in the form of Schr\"odinger equation with
the axionlike potential $W$.

Eq.~(\ref{SG1}) is invariant under the global transformation
$\chi(\xi)\mapsto\chi(\xi)+2\pi n$, $n\in\mathbb{Z}$, while
Eq.~(\ref{feq1}) is not. Its general solution is easily derived
by integration, which is carried out in numerous works
(see, for instance, Sec.~5.3 in \cite{Man83}).
For physical reasons, we write down and exploit the stationary anti-kink solution
\begin{equation}\label{SGsol}
\chi_{\rm ak}(\xi)=4\arctan{\rme^{-\xi}},\quad
\xi\geq0.
\end{equation}
We also consider    
the solution of the form $\chi_{\rm ak}(\xi-L)$ with
an arbitrary constant $L$. Altogether these solutions at dimensionless
energy $\varepsilon=0$ describe the ground state and, moreover, devoid any
nodes, in contrast to the oscillating solutions in
Sec.~\ref{S2}. According to (\ref{SG1}), they determine the potential
$W$ in terms of the coordinate $\xi$.

It is clear that the solution (\ref{SGsol}) can be also obtained
from Eq.~(\ref{SG1}) with the potential $W$ depending directly on $\xi$.
Using the auxiliary formula
\begin{equation}\label{sin4}
\sin{4z}=4\,\frac{1-\tan^2{z}}{\left(1+\tan^2{z}\right)^2}\,\tan{z},
\end{equation}
one arrives at $W$ of the form
\begin{equation}\label{Wak}
W_{\rm ak}(\xi)=\frac{\tanh{\xi}}{2\cosh{\xi} \arctan{\rme^{-\xi}}}.
\end{equation}

Note that, replacing $\chi$ with $4\arctan{\varphi}$, the sine-Gordon
potential reduces to the form $\sin{\chi}=4\varphi \cos_{\mu=1}{\varphi}$
accordingly to (\ref{sin4}), where $\cos_{\mu}{z}$ is the $\mu$-deformed
cosine-function~\cite{GN22p,GKN23} taken at $\mu=1$. Moreover, $\cos_{\mu=1}{\xi}$
is used in \cite{GN22p} to simulate the potential of two coupled axions
at the quantum mechanical level.

To reproduce (\ref{SGsol}) by solving Eq.~(\ref{SG1}), we have 
to take $\chi_{\rm ak}(0)=\pi$ and $\chi_{\rm ak}^\prime(0)=-2$.
 The same approach relates the potential $W_{\rm ak}(\xi-L)$ with  the solution
$\chi_{\rm ak}(\xi-L)$. 
At $\xi=0$, we set $\chi_{\rm ak}=4\arctan{\rme^{L}}$
and $\chi_{\rm ak}^\prime=-2/\cosh{L}$.

\begin{figure}[htbp]
\includegraphics[width=6.2cm,angle=0]{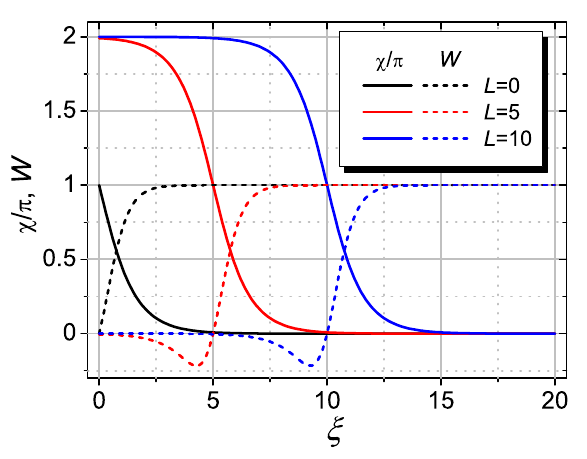} 

\vspace*{-1.8mm}
\caption{\label{XW}Antikink solution $\chi(\xi-L)$ (solid lines) and
corresponding potential $W(\xi-L)$ (dashed lines) for different $L$.}
\end{figure}

The behavior of $\chi_{\rm ak}(\xi-L)$ and $W_{\rm ak}(\xi-L)$ is shown in Fig~\ref{XW}.
For fixed $L$, we see that $\chi_{\rm ak}\to0$ and $W_{\rm ak}\to1$ for $\xi\to\infty$
as long as $\chi_{\rm ak}$ reaches its maximum at $\xi=0$. For large $L$,
the value of $\chi_{\rm ak}(\xi-L)$ at $\xi=0$ tends to $2\pi$, which is similar
to the single-phase solution in Fig.~\ref{chi}(a), colored in black.
We can deduce that the profile of $\chi_{\rm ak}$ qualitatively depicts the DM halo core
due to the potential $W_{\rm ak}$, which results in an infinite scattering
length $a$ in the Born approximation and forms a closed scattering channel
for particles with zero total energy $\varepsilon$ (or $\nu$), which are unable
to overcome the potential barrier/domain wall.

Although the gravitational interaction of a large number of relatively
fast particles modifies this barrier [cf. Fig.~\ref{chi}(b)],
the mechanism of injection of a slow particle into a closed channel
is of special interest.
One possibility which we further explore is the transfer of particle between
different scattering channels using the Feshbach resonance
stimulated by an additional impact.

Let us calculate the integral over the entire (one-dimensional) space:
\begin{eqnarray}
N(L)&\equiv&\int_0^\infty \chi^2_{\rm ak}(\xi-L)\,\rmd\xi\label{NumL}\\
&=&4\int_0^{\alpha} \frac{z^2}{\sin{z}}\,\rmd z;\qquad
\alpha=2\arctan{\rme^L}.
\label{NumL2}
\end{eqnarray}
Note that $2\alpha=\chi_{\rm ak}(\xi-L)$ at $\xi=0$, that is
the (maximal) value of axion field at the origin.

\begin{figure}[htbp]
\includegraphics[width=6.2cm,angle=0]{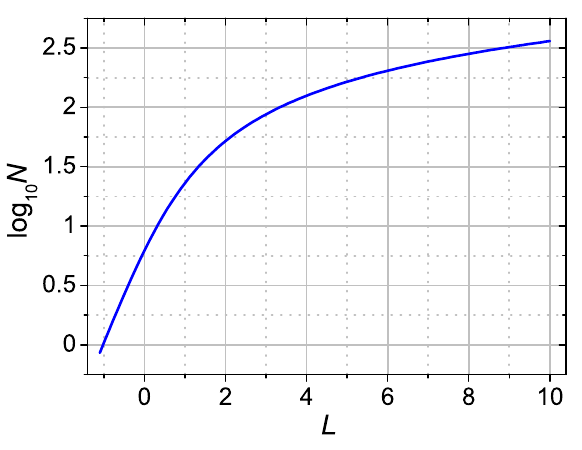} 

\vspace*{-1.8mm}
\caption{\label{NL}Normalization~(\ref{NumL}) associated with
the number of particles as a function of length parameter $L$.}
\end{figure}

On integrating, the result is presented in differing forms:
\begin{eqnarray}
N&=&2\alpha^2+4\sum\limits_{n=1}^\infty (-1)^{n+1}
\frac{2^{2n-1}-1}{(n+1) (2n)!}\,B_{2n} \alpha^{2n+2}\nonumber\\
&=&2\rme^{\rmi\alpha}\,\left[
\Phi\left(\rme^{2\rmi\alpha},3,\frac{1}{2}\right)
-2\rmi\alpha\,\Phi\left(\rme^{2\rmi\alpha},2,\frac{1}{2}\right)\right]\nonumber\\
&&+4\alpha^2 \left(\ln{\tan{\frac{\alpha}{2}}}-\rmi\frac{\pi}{2}\right)-14\, \zeta(3),
\end{eqnarray}
where $B_{2n}$ is the Bernoulli number; $\Phi(z,s,a)$ is
the Lerch transcendent; $\zeta(3)=1.20205...$ is the particular
value of Riemann zeta-function; at last, $\rmi=\sqrt{-1}$.
The first expression is valid for $\alpha<\pi$ and given by
Eq.~(1.5.44.1) in \cite{PBM}.
Behavior of $N(L)$ is also shown in Fig.~\ref{NL}.

By construction, $N(L)$ is related with the number of particles in
nonlinear problem (\ref{SG1}). This can be useful, for instance,
to control the effect of gravitation, as mentioned above.
It is obvious that $N(L)$ is distinct from the (anti)kink topological
charge~\cite{Man83}.

On the other hand, when formulating the linear Schr\"odinger equation
with potential $W_{\rm ak}(\xi-L)$, we can use $N(L)$ to normalize
$\chi_{\rm ak}(\xi-L)$, the fact that will be applied below
in the quantum mechanical setting.

Thus, we summarize that, neglecting gravity when considering a small
number of axions, their (ground) bound state is still revealed even
in the one-dimensional case, which nevertheless retains main physical
properties of the model we need and only simplifies the mathematical
description. Besides, the model (field) equation is reformulated in
an equivalent quantum-mechanical form through introducing an effective
potential depending on space.

\section{\label{S4}Feshbach Resonance}

We would like to consider in more detail the mechanism of the transition
of a DM particle into the bound state described above, noting that the presence
of one good potential is apparently not enough for this. We appeal to
resonance scattering, which implies the existence of an isolated (quasi-discrete)
energy level.

Our approach inherits the ideas of
Feshbach resonance~\cite{Fesh62}, which
uses at least two channels of scattering, one open and one closed channels
with distinct Hamiltonians. Coupling these channels enables to create
an isolated level and bind the scattered particle. This approach seems
to be appropriate, since it is difficult to directly transfer a zero-energy
particle through the domain wall into the trap of sine-Gordon potential.

Let us assume that a spinless particle is able to jump between open and
closed channels. When the total energy exceeds the open channel threshold
($E=0$), the open channel
becomes both an incoming and an outgoing channel. The Feshbach resonance occurs
when the energy of the bound state in the closed channel is close to the threshold
of the open channel. Due to the coupling of the channels, the unperturbed bound
state of the closed channel becomes dressed. This dressed state is treated as a
quasi-bound state of the scattered particle. The scattered particle temporarily
passes into the quasi-bound state with positive energy and returns to the open
channel after a typical time delay $\tau=2/\Gamma$, determined by the decay
width $\Gamma$ of the quasi-bound state. We associate this state of two
interacting particles with a composite (dimer).

\subsection{\label{S4A}Preliminaries and Model Formulation}

Formulating our model on the base of stationary Schr\"odinger equation,
we consider in fact a slowly moving particle which is assumed to be
in one of the two channels. One channel is {\it closed} and is described in
the absence of external interaction by the sine-Gordon equation in the
ground state with energy $E^{(1)}=0$, as mentioned earlier in Sec.~\ref{Sec3}.
An {\it open} (entrance) channel, the existence of which we assume,
corresponds to elastic scattering due to another interaction, where
a wave/particle initially has low energy $E^{(2)}=k^2>0$ determined by
momentum~$k$, which we also treat as the relative momentum of the pair of
interacting particles. Then, there is an energy gap between these channels
$Q=E^{(2)}-E^{(1)}>0$, which is further affected by external impact. Thus,
such a model involves three different interactions, and $E^{(1)}$ and $E^{(2)}$
are not energy levels of the same Hamiltonian. For this reason, the interaction
parameters must be tuned to obtain the desired effect of resonant transition.

We describe the motions of a particle in two channels, adopting
the matrix representation~\cite{Yam93}:
\begin{equation}\label{Meq1}
\mathbb{H} X=E X,\qquad
\mathbb{H}=
\left(
\begin{array}{cc}
H_{\rm bs} & \Omega \\
\Omega^\dag & H_{\rm wv}
\end{array}
\right),
\end{equation}
where $H_{\rm bs}$ and $H_{\rm wv}$ are the Hamiltonians of the bound
state (in closed channel) and the scattered wave (in open channel),
respectively. The coupling between channels is represented by $\Omega$
and is associated with extra force, which is turned on and starts to
act {\it after} fixing the gap $Q$. In other words, $E^{(1)}$ and
$E^{(2)}$ are given at $\Omega=0$, while energy $E$ is determined
by switching $\Omega$.

We account for the energy gap $Q$ in Eq.~(\ref{Meq1}) by defining
$H_{\rm bs}=H_{\rm ak}+Q$, where
\begin{equation}
H_{\rm ak}=-\frac{\rmd^2}{\rmd\xi^2}+W_{\rm ak}(\xi-L)
\end{equation}
with the potential $W_{\rm ak}(\xi)$ from Eq.~(\ref{Wak}), and $W_{\rm ak}(0)=0$.

Hence, we use the dependence on parameter $L>0$, which plays an
important role in further constructions.

In a sense, the system is doubly degenerate at $\Omega=0$ due to existing
two independent wave functions for the same eigenvalue $E$:
\begin{equation}\label{P0}
X^{(1)}=\left(
\begin{array}{c}
\chi^{(1)}\\
0
\end{array}
\right),\qquad
X^{(2)}=
\left(
\begin{array}{c}
0\\
\chi^{(2)}
\end{array}
\right),
\end{equation}
which are evidently orthogonal in this representation.
We identify $\chi^{(1)}$ with $\chi_{\rm ak}(\xi-L)$
from Eq.~(\ref{SGsol}). In principle, we need to write
$X=X^{(1)} \cos{\alpha}+X^{(2)} \sin{\alpha}$ with some $\alpha$
in order to normalize the total wave function $X$ with respect
to the matrix representation.

As shown in Fig.~\ref{XW}, the spatial interval $\xi\in[0;L]$
is most significant for the manifestation of a bound state. Therefore,
essential processes should be related with this region, which defines
the {\it resonance zone}. For this reason, we concentrate there on
the external force, which is parametrized by $\omega$ as
\begin{equation}\label{exi}
\Omega(\xi)=-\omega^2\,\theta(L-\xi),\qquad \Omega^\dag=\Omega.
\end{equation}
Here $\theta$ is the Heaviside step-function.

Similarly, we define the interaction in the open channel by
the square-well potential:
\begin{equation}\label{Vi}
V_{\rm sq}(\xi)=-V\,\theta(L-\xi),\quad
V>0.
\end{equation}
The strength $V$ along with $\omega^2$ are the variable parameters of the model.

Since we are studying the mechanism of the emergence of resonance and
two-particle composite, the refinement of the nature and form of these
extra interactions remains for further consideration. Here we only use
their simplest version 
and discuss their origin after the computations performed.

By construction, all spatial functions in such a model are divided into
two components belonging either to the interval $\xi\in[0;L]$ or to the interval
$\xi\in[L;\infty)$, which we label by ``$<$'' and ``$>$'' relative 
to the separating point $\xi=L$. Then, the wave functions for the channels
are numbered by $\alpha=1,2$ and decomposed as
\begin{equation}
\chi^{(\alpha)}(\xi)=\theta(L-\xi)\,\chi^{(\alpha)}_{<}(\xi)+\theta(\xi-L)\,\chi^{(\alpha)}_{>}(\xi).
\end{equation}

We connect the functions at separating point $\xi=L$ by the matching 
condition
\begin{equation}\label{rBC}
\left.\frac{\rmd}{\rmd\xi}\,\ln{\chi^{(\alpha)}_{<}(\xi)}\right|_{\xi=L}=
\left.\frac{\rmd}{\rmd\xi}\,\ln{\chi^{(\alpha)}_{>}(\xi)}\right|_{\xi=L},
\end{equation}
to guarantee the equality of derivatives and proportionality of
the functions in the left and right sides of (\ref{rBC}).

Before proceeding further, we recall the known results for the open channel
($\alpha=2$) in the absence of coupling $\Omega$. The scattering
characteristics result from the equation
\begin{equation}\label{sq1}
H_{\rm wv} \chi^{(2)}\equiv\left(-\frac{\rmd^2}{\rmd\xi^2}+V_{\rm sq}(\xi)\right) \chi^{(2)}
=E \chi^{(2)}.
\end{equation}

Solution to Eq.~(\ref{sq1}) is given as
\begin{eqnarray}
&&\hspace*{-3mm}
\chi^{(2)}(\xi)=\theta(L-\xi)\,\chi^{(2)}_{<}(\xi)+\theta(\xi-L)\,\chi^{(2)}_{>}(\xi),\label{ch2}\\
&&\hspace*{-3mm}
\chi^{(2)}_{<}(\xi)=\sin{K\xi},\quad\
\chi^{(2)}_{>}(\xi)=\sin{KL}\,\frac{\sin{(k\xi+\delta)}}{\sin{(kL+\delta)}},
\label{unp2}
\end{eqnarray}
where $K=\sqrt{E+V}$ and $k=\sqrt{E}$, that is, $E=k^2$.
Note that $\chi^{(2)}_{>}$ behaves as $\exp{(-\sqrt{|E|}\xi)}$ when $E<0$.

The phase shift $\delta$ is derived from the relation (\ref{rBC}):
\begin{equation}\label{screl1}
K \cot{(KL)}=k \cot{(kL+\delta)}.
\end{equation}

Then, computing the scattering length as
\begin{equation}\label{adef}
a=-\lim\limits_{k\to0}\frac{\tan{\delta(k)}}{k},
\end{equation}
one finds its expression for potential~$V_{\rm sq}(\xi)$:
\begin{equation}\label{aV}
a_V=L\,\left(1-\frac{\tan{(\sqrt{V}L)}}{\sqrt{V}L}\right).
\end{equation}

Note that $a_V$ demonstrates discontinuous behavior for
$\sqrt{V}L=(2n-1)\pi/2$ and $n\in\mathbb{N}$. We omit
the detailed consideration of this zero-energy resonance.

Let us emphasize the essential difference between the physical
consequences of the zero and divergent scattering lengths $a_V$
in one and three dimensions, despite the formal similarity of
the presented expressions to the three-dimensional case~\cite{LL}.
While the vanishing $a_V$ means complete transparency in three dimensions,
the opposite effect occurs in one dimension: the reflection coefficient
becomes equal to unity, which leads to complete opacity. Transparency
in one dimension is achieved when $a_V$ diverges. Nevertheless,
the divergence of the scattering length reveals a zero-energy bound state
in both three-dimensional and one-dimensional cases~\cite{Tar13}.

In one dimension (see \cite{E65}), the scattering matrix
$S=\rme^{2\rmi\delta}$ and amplitude $f$ are
\begin{equation}\label{scam1}
S=\rme^{-2\rmi kL}\,\frac{K \cot{(KL)}+\rmi k}{K \cot{(KL)}-\rmi k},\quad
f=\frac{1}{2}\left(\rme^{2\rmi\delta}-1\right).
\end{equation}
These formulas tell us how to extract scattering data
in the open channel.

\subsection{\label{S4B}Two-Channel Quantum Mechanics}

Thus, we admit a single bound state in the closed channel and
a continuum of waves with momentum $k$ in the open channel.
Taking into account the complexity of the problem involving
anti-kink potential, we intend to analytically describe
the Feshbach resonance between the channels in the first approximation.

As mentioned above, the initial set of equations in entire space is
\begin{eqnarray}
(H_{\rm ak}+Q-E)\chi^{(1)}+\Omega\chi^{(2)}&=&0,\label{e1-0}\\
(H_{\rm wv}-E)\chi^{(2)}+\Omega^\dag \chi^{(1)}&=&0.\label{e2-0}
\end{eqnarray}

For convenience, we will use the bra- and ket-vectors to simplify
the notation of matrix elements.

In the first approximation, we put~\cite{Joa75}:
\begin{equation}\label{anz1}
|\chi^{(1)}\rangle=\frac{\lambda}{N(L)}\,|\chi_{\rm ak}\rangle,\quad
\langle\xi|\chi_{\rm ak}\rangle=\chi_{\rm ak}(\xi-L),
\end{equation}
where $\lambda$ is a complex constant which should be found; $N(L)$ is given by Eq.~(\ref{NumL}).

Acting by $\langle\chi_{\rm ak}|$ on Eq.~(\ref{e1-0}), one has
\begin{equation}
\lambda=\frac{\langle\chi_{\rm ak}|\Omega|\chi^{(2)}_{<}\rangle}{E-Q},
\end{equation}
where it has been used that $H_{\rm ak}|\chi_{\rm ak}\rangle=0$, the normalization
$\langle\chi_{\rm ak}|\chi_{\rm ak}\rangle=N(L)$, and the equality
$\langle\chi_{\rm ak}|\Omega|\chi^{(2)}\rangle=\langle\chi_{\rm ak}|\Omega|\chi^{(2)}_{<}\rangle$
due to the form of $\Omega(\xi)$. At this stage, the coefficient $\lambda$
still depends on the unknown function $\chi^{(2)}_{<}$.

Introducing the auxiliary Hamiltonian
\begin{equation}
H_2=-\frac{\rmd^2}{\rmd\xi^2}-K^2,\qquad
K^2=E+V,
\end{equation}
which is defined in the region $\xi\in[0;L]$,
the equations for the open channel take the form
\begin{eqnarray}
H_2|\chi^{(2)}_{<}\rangle+\Omega^\dag|\chi^{(1)}\rangle&=&0,\label{e2-p1}\\
-\frac{\rmd^2\chi^{(2)}_{>}}{\rmd\xi^2}-k^2\chi^{(2)}_{>}&=&0.\label{e2-p2}
\end{eqnarray}

Solution to Eq.~(\ref{e2-p1}) can be written as
\begin{eqnarray}
|\chi^{(2)}_{<}\rangle&=&|\tau_0\rangle-G^{(+)}_2\Omega^\dag|\chi^{(1)}\rangle
\nonumber\\
&=&|\tau_0\rangle-\frac{\lambda}{N(L)}\, G^{(+)}_2\Omega^\dag|\chi_{\rm ak}\rangle,
\label{e2-3}
\end{eqnarray}
where unperturbed wave function $\tau_0(\xi)=\langle\xi|\tau_0\rangle$
coincides with $\chi^{(2)}_{<}(\xi)$ from Eq.~(\ref{unp2}) and is such that
\begin{equation}
H_2\,\tau_0(\xi)=0,\qquad
\tau_0(\xi)=\sin{K\xi}.
\end{equation}

The Hamiltonian $H_2$ determines also the Green's operator
$G^{(+)}_2=(H_2-\rmi\epsilon)^{-1}$ that contains the shifted
energy $E+\rmi\epsilon$ at $\epsilon\to0$.
The corresponding Green's function is
\begin{eqnarray}
&&\hspace*{-4.2mm}
G^{(+)}_2(\xi,\zeta;K)=\frac{\sin{K\xi_{1}} \cos{K\xi_{2}}}{K}
+\rmi\, \frac{\sin{K\xi} \sin{K\zeta}}{K},\\
&&\hspace*{-4mm}
\xi_{1}={\rm min}(\xi,\zeta),\qquad \xi_{2}={\rm max}(\xi,\zeta),
\nonumber
\end{eqnarray}
which serves for finding the outgoing wave under the boundary
condition $G^{(+)}_2(0,\zeta;K)=0$.

To express $\lambda$ in terms of the known solutions $\tau_0$
and $\chi_{\rm ak}$, let us operate by $\langle\chi_{\rm ak}|\Omega$
on Eq.~(\ref{e2-3}). Then we obtain
\begin{equation}
\lambda=\frac{\langle\chi_{\rm ak}|\Omega|\tau_0\rangle}
{E-Q+N^{-1}(L)\langle\chi_{\rm ak}|\Omega G^{(+)}_2\Omega^\dag|\chi_{\rm ak}\rangle}.
\end{equation}
The condition of vanishing of the denominator reveals
the isolated (quasi-discrete) energy level of the dressed
state~\cite{Fesh62,Joa75}.

Having introduced the notations
\begin{eqnarray}
\omega^4\Delta_L(K)&=&N^{-1}(L)\,{\rm Re}\,\langle\chi_{\rm ak}|\Omega G^{(+)}_2\Omega^\dag|\chi_{\rm ak}\rangle,
\label{Del}\\
\omega^4\gamma_L(K)&=&N^{-1}(L)\,{\rm Im}\,\langle\chi_{\rm ak}|\Omega G^{(+)}_2\Omega^\dag|\chi_{\rm ak}\rangle,
\label{gam}
\end{eqnarray}
let us sketch how this works for a fixed $Q>0$.
We first imagine the situation when $V\gg E$ for $E\to0$,
and the denominator of $\lambda$ vanishes at some complex value of
the energy $E^\prime-\rmi\Gamma^\prime/2$, thereby making the magnitude
of the wave functions extremely large. The resonance energy
$E=E^\prime$ and decay width $\Gamma^\prime$ may be simply determined:
\begin{equation}\label{Rch}
E^\prime=Q-\omega^4\Delta_L(\sqrt{V}),\qquad
\Gamma^\prime=2\omega^4\gamma_L(\sqrt{V}).
\end{equation}
For relatively small $\omega^2$ and positive $\Delta_L(\sqrt{V})$,
we can achieve that $Q>E^\prime>0$ due to two additional interactions.
Besides, for positive $\gamma_L(\sqrt{V})$, the lifetime of particle
in such a state is $\tau=2/\Gamma^\prime$, in dimensionless units.

In general case we write
\begin{equation}
\lambda=\frac{\langle\chi_{\rm ak}|\Omega|\tau_0\rangle}
{E-Q+\omega^4\Delta_L(K)+\rmi \omega^4\gamma_L(K)}.
\end{equation}
Consider the overlap integral that defines $\langle\chi_{\rm ak}|\Omega|\tau_0\rangle$:
\begin{equation}\label{Bdef}
B_L(K)\equiv 4\int_0^L \arctan{\rme^{L-\xi}}\,\sin{K\xi}\,\rmd\xi.
\end{equation}
It can be transformed to the form
\begin{eqnarray}
B_L(K)&=&\frac{1}{2\rmi}\,[\phi_L(L,\rmi K)-\phi_L(L,-\rmi K)]\nonumber\\
&&-\frac{1}{2\rmi}\,[\phi_L(0,\rmi K)-\phi_L(0,-\rmi K)].
\end{eqnarray}
Using the Lerch transcendent $\Phi(z,s,a)$, we introduce
\begin{eqnarray}
\phi_L(\xi,a)&\equiv&\left.\frac{\rme^{a\xi}}{a}\,\right[4\arctan{\rme^{L-\xi}}\nonumber\\
&&\left.-2\,\rme^{L-\xi}\,\Phi\left(-\rme^{2(L-\xi)},1,\frac{1-a}{2}\right)\right].
\end{eqnarray}
This function is such that
\begin{equation}
\partial_\xi\phi_L(\xi,a)=\rme^{a\xi}\chi_{\rm ak}(\xi-L).
\end{equation}
Thus, we have $\langle\chi_{\rm ak}|\Omega|\tau_0\rangle=-\omega^2 B_L(K)$.

\begin{figure}
\begin{center}
\includegraphics[width=5.6cm]{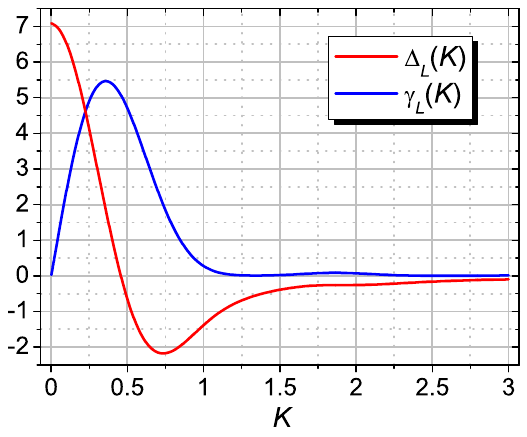} 
\end{center}
\vspace*{-5mm}
\caption{\label{fig5}The functions $\Delta_L(K)$ and $\gamma_L(K)$ which determine
$\langle\chi_{\rm ak}|\Omega G^{(+)}_2\Omega^\dag|\chi_{\rm ak}\rangle$ at $L=5$.
There is $K_{\Delta}\simeq0.4563077$ such that $\Delta_L(K_{\Delta})=0$.} 
\end{figure}

As seen above, the first correction to the wave function $\chi^{(2)}_{<}(x)$
is determined by $G^{(+)}_2\Omega^\dag|\chi_{\rm ak}\rangle$. To find the
complex function
$\langle\xi|G^{(+)}_2\Omega^\dag|\chi_{\rm ak}\rangle=-\omega^2\tau_1(\xi)$,
we turn to the solution of the inhomogeneous equation
\begin{eqnarray}
&&H_2\,\tau_1(\xi)=\chi_{\rm ak}(\xi-L),\\
&&\tau_1(\xi)=\int_0^L G^{(+)}_2(\xi,\zeta;K)\,\chi_{\rm ak}(\zeta-L)\,\rmd\zeta.
\end{eqnarray}

On computing, the summands of
$\tau_1(\xi)=\tau^{\rm R}_1(\xi)+\rmi\,\tau^{\rm I}_1(\xi)$ are written as
\begin{eqnarray}
\tau^{\rm R}_1(\xi)&=&\frac{\rme^{-\rmi K\xi}}{2\rmi K}\,\phi_L(\xi,\rmi K)
-\frac{\rme^{\rmi K\xi}}{2\rmi K}\,\phi_L(\xi,-\rmi K)\nonumber\\
&&+[\phi_L(L,\rmi K)+\phi_L(L,-\rmi K)]\,\frac{\sin{(K\xi)}}{2K}\nonumber\\
&&+\rmi\,[\phi_L(0,\rmi K)-\phi_L(0,-\rmi K)]\,\frac{\cos{(K\xi)}}{2K},\\
\tau^{\rm I}_1(\xi)&=&B_L(K)\,\frac{\sin{(K\xi)}}{K}.
\end{eqnarray}

Taking into account the form of $\tau^{\rm I}_1(\xi)$ and the definition
of $B_L(K)$, we specify the function $\gamma_L(K)$ [see Eq.~(\ref{gam})]:
\begin{equation}\label{gamL}
\gamma_L(K)=\frac{B_L^2(K)}{K\,N(L)}\geq0.
\end{equation}

At the same time, the real part $\tau^{\rm R}_1(\xi)$ determines
also the deviation $\Delta_L(K)$ so that
\begin{equation}
\Delta_L(K)=N^{-1}(L) \int_0^L \chi_{\rm ak}(\xi-L)\,\tau^{\rm R}_1(\xi)\,\rmd\xi.
\end{equation}
This integral is not simple to be calculated analytically, instead we present
the numerical result in Fig.~\ref{fig5}.

Let us note the similarity of the behavior of $\gamma_L(K)$ and $\Delta_L(K)$
with analogous functions from Ref.~\cite{GN22p}, which are calculated for
another potential in three dimensions. {\it This means that the number of spatial
dimensions does not affect main physical aspect of the problem}.
By combining, the closed channel wave function in the first approximation reads
\begin{equation}\label{ss1}
\chi^{(1)}(\xi)=-\frac{\omega^2}{D_L(K)}\,\sqrt{\frac{K\gamma_L(K)}{N(L)}}\,\chi_{\rm ak}(\xi-L),
\end{equation}
where we have used the notation $D_L(K)$ introduced as
\begin{eqnarray}
D_L(K)&=&K^2-V-Q+\omega^4\Delta_L(K)+\rmi \omega^4\gamma_L(K)
\nonumber\\
&=&D_L^{\rm R}(K)+\rmi D_L^{\rm I}(K).\label{deno}
\end{eqnarray}

Solution~(\ref{ss1}) vanishes at $\omega=0$ and describes a short-lived 
state, that is seen by restoring for a moment the time dependence due to
decaying factor $\exp{(-\rmi Et)}$ determined by the complex
$E=Q-\omega^4\Delta_L(K)-\rmi\omega^4\gamma_L(K)$.

Combining the terms with $\tau_0(\xi)$ and $\tau^{\rm I}_1(\xi)$ due to
proportionality $\tau^{\rm I}_1(\xi)\propto\tau_0(\xi)$, we present
the open channel wave function in the resonance zone as
\begin{equation}
\chi^{(2)}_{<}(\xi)=\frac{D_L^{\rm R}(K)}{D_L(K)}\,\tau_0(\xi)
-\frac{\omega^4}{D_L(K)}\,\sqrt{\frac{K\gamma_L(K)}{N(L)}}\,\tau^{{\rm R}}_{1}(\xi).
\label{ss2<}
\end{equation}

To find the phase shift $\delta$ for the function $\chi^{(2)}_{>}(\xi)$ namely
\begin{eqnarray}
\chi^{(2)}_{>}(\xi)&=&\chi^{(2)}_{<}(L)\,\frac{\sin{(k\xi+\delta)}}{\sin{(kL+\delta)}};
\label{ss2>}\\
\chi^{(2)}_{<}(L)&=&\frac{D_L^{\rm R}(K) \sin{KL}-D_L^{\rm I}(K) \cos{KL}}{D_L(K)},\nonumber
\end{eqnarray}
we appeal to the matching condition (\ref{rBC}).  We obtain
\begin{equation}
K\,\frac{D_L^{\rm R}(K) \cos{KL}+D_L^{\rm I}(K) \sin{KL}}{D_L^{\rm R}(K) \sin{KL}-D_L^{\rm I}(K) \cos{KL}}
=k \cot{(kL+\delta)}.\nonumber
\end{equation}
This relation can be re-written as
\begin{eqnarray}
&&K \cot{(KL-\delta_{\rm rs})}=k \cot{(kL+\delta)},\\
&&\delta_{\rm rs}(K)=\arctan{\frac{D_L^{\rm I}(K)}{D_L^{\rm R}(K)}},
\end{eqnarray}
where $\delta_{\rm rs}$ is the phase shift caused by interactions in
the resonance zone. At $\delta_{\rm rs}\equiv0$, only potential
scattering with $V_{\rm sq}(x)$ remains in the open channel.

Now it is easy to extract the total phase shift:
%
\begin{eqnarray}
\delta&=&-kL+\arctan{\left[\frac{k}{K}\,\tan{(KL-\delta_{\rm rs})}\right]}\label{phs1}\\
&\simeq&k\left[\frac{\tan{(KL-\delta_{\rm rs})}}{K}-L\right],\nonumber
\end{eqnarray}
where the second expression is used for $k\to0$.

\subsection{\label{S4C}Feshbach Phenomenon}

Let us dwell on the effects at zero energy and momentum $k$.
It is reasonable to study the properties of the scattering length $a$
of the particles in the open channel, having got the phase shift $\delta$
in (\ref{phs1}) and using the Eq.~(\ref{adef}). For our purposes,
we represent $a$ in the form
\begin{equation}\label{scl1}
a(\omega^2)=a_{\rm bg}\,\left(1+\frac{{\cal D}}{\omega^4-\omega^4_c}\right).
\end{equation}
In this formula, we have explicitly taken into account the dependence
on the magnitude of the external influence $\omega^2$ [see (\ref{exi})]
and shown the presence of critical value $\omega^2_c$:
\begin{equation}\label{omc}
\omega^2_c=\frac{\sqrt{Q}}{\sqrt{\Delta_L(K_V)+\gamma_L(K_V) \tan{(K_VL)}}};
\quad K_V\equiv\sqrt{V}.
\end{equation}
This is determined by the energy gap $Q>0$ between the two channels,
which actually coincides with the kinetic energy of the incident particle
outside the resonance zone at $\xi>L$. Avoiding here the zero-energy
resonances in the open channel at $K_VL=(2n-1)\pi/2$ for $n\in\mathbb{N}$,
we require $0<\Delta_L(K_V)+\gamma_L(K_V) \tan{(K_VL)}<\infty$ to ensure
a real value of $\omega^2_c$.

The remaining characteristics are given as
\begin{eqnarray}
&&a_{\rm bg}=a_V+\ell, \qquad {\cal D}=\omega^4_c\,\frac{\ell}{a_{\rm bg}},\\
&&\ell=\frac{1}{K_V}\,\frac{\gamma_L(K_V)[1+\tan^2{(K_VL)}]}
{\Delta_L(K_V)+\gamma_L(K_V) \tan{(K_VL)}}>0,
\end{eqnarray}
where $a_V$ is defined by (\ref{aV}), while $a_{\rm bg}$ is
the so-called background scattering length.
Note that the dependence of $a$ on $\omega^2$ vanishes at $Q=0$ ($\omega^2_c=0$)
so that $a=a_V+\ell$.

At $V\to0$, we obtain $a_V=0$, and
\begin{eqnarray}
a&=&\ell_0\,\frac{\omega^4}{\omega^4-\omega^4_c},\qquad
\omega^4_c=\frac{Q}{\Delta_L(0)},\label{a22}\\
\ell_0&=&\frac{1}{N(L)\,\Delta_L(0)} \lim\limits_{K\to0} \left(\frac{B_L(K)}{K}\right)^2,
\end{eqnarray}
where $B_L(K)$ is defined by Eq.~(\ref{Bdef}), and Eq.~(\ref{gamL}) is used.

Thus, we gain  
the relation $a=\ell_0(L)\propto L$ when both $Q=0$ and $V=0$.

\begin{figure}
\begin{center}
\includegraphics[width=5.8cm]{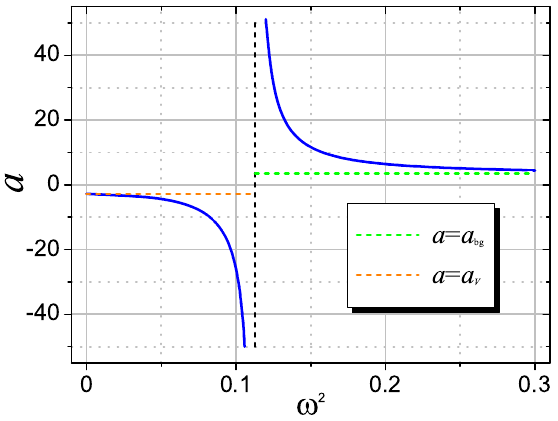} 
\end{center}
\vspace*{-5mm}
\caption{\label{fig6}Scattering length $a$ as the function of
external impact $\omega^2$. The dashed straight lines are asymptotics.
Black line corresponds to $\omega^2=\omega^2_c$, while green and orange
lines are for $a=a_{\rm bg}$ and $a=a_{V}$, respectively.} 
\end{figure}

In general, the Feshbach phenomenon which is in our focus,  
results from the {\it dependence of the scattering length $a$ on the interaction parameters},
that allows us to detect bound states by the divergence of $a$.
Our model assumes that the change in $a$ depends on $Q$ (defining $\omega^2_c$)
and on $\omega^2$, while $V>0$ is given.

Analyzing  Eq.~(\ref{scl1}), let us fix $L=5$, $K_V=0.2$ that provides
$K_V<K_{\Delta}$ and $\Delta_L(K_V)>0$ (see Fig.~\ref{fig5}), and
$Q=0.15$. It leads to the following characteristics in dimensionless
units: $a_V\simeq-2.787039$,
$a_{\rm bg}\simeq3.467758$, $\omega^2_c\simeq0.1128185$, and
${\cal D}\simeq0.0229575$. The dependence of $a$ on the external
impact strength $\omega^2$ is shown in Fig.~\ref{fig6}, which confirms
that the bound state does indeed occur at $\omega^2=\omega^2_c$.
Critical value $\omega^2_c$ determines the threshold of the production of
shallow dimers at $a(\omega^2)\gg L$ with the binding energy~\cite{BrH06}
\begin{equation}\label{Ebi}
E_{\rm bind}=-\varkappa^2\propto-\frac{1}{a^2(\omega^2)}.
\end{equation}

This follows from considering the case of small negative energy $E\to0^{-}$,
when $\chi^{(2)}_{>}\propto\exp{[-\varkappa(\xi-L)]}$ for $\xi>L$, and
Eq.~(\ref{rBC}) yields
\begin{equation}
-\varkappa=\lim\limits_{k\to0}\left.\frac{\rmd}{\rmd\xi}\ln{\chi^{(2)}_{<}(\xi)}\right|_{\xi=L}
=-\frac{1}{a(\omega^2)-L}.\label{kabi}
\end{equation}

Thus, the Feshbach phenomenon justifies the need for a large scattering
length in the formation of composites (of at least two particles), as
predicted in Ref.~\cite{GN22} using phenomenological approaches.
This is due to the fact that the Feshbach phenomenon as a zero-energy
effect is valid in a different number of spatial dimensions,
although there are distinct physical implications of zero and diverging
scattering lengths in one and three dimensions~\cite{Tar13}.

In principle, the energy gap $Q$ can be maintained by a spatially
homogeneous interaction that induces the energy predominance of one
configuration of the system of particles over another.
In this regard, we mention experiments with alkali atoms, the energy
configurations of which are determined by the spin and the applied
magnetic field. Therefore, there, the Feshbach phenomenon is related
with a resonant transition between configurations with a change of
the magnetic field~\cite{Grimm10,BrH06}.

\subsection{\label{S4D}Resonance Scattering}

To reveal the newly formed bound state (of two axions),
we also investigate the resonance scattering of an incident particle
with a nonzero energy $E=k^2$.

The scattering matrix element $S=\rme^{2\rmi\delta}$ for the open channel
[cf. Eq.~(\ref{scam1})] is
\begin{equation}
S=\rme^{-2\rmi kL}\,\frac{K \cot{(KL-\delta_{\rm rs})}+\rmi k}
{K \cot{(KL-\delta_{\rm rs})}-\rmi k}.
\end{equation}
Denoting its denominator as
\begin{equation}
F(k)=K \cot{(KL-\delta_{\rm rs})}-\rmi k;\quad
K=\sqrt{k^2+V},
\end{equation}
the condition $F(k)=0$ determines the pole of $S$ and the resonance point
(although, not every pole of $S$ is related with the compound system
existence).

Usually, a resonance is observed in narrow region of energy~$E$,
which covers the resonant value $E_{\rm res}=E_0-\rmi \Gamma_0/2$
with some $E_0>0$ and $\Gamma_0>0$. Positivity of $E_0$ makes this
level unstable, whose lifetime is $\tau=2/\Gamma_0$, in dimensionless units.

Here, we find $E_{\rm res}=K^2_{\rm res}-V$ for given $Q$, $V$,
$\omega$, and $L$ by solving the equation $F(\sqrt{K^2_{\rm res}-V})=0$
in an appropriate form. After identical transformation is performed,
the following equation should be solved numerically at $\omega^4\ll1$ by
using an iterative procedure with the initial value $K=\sqrt{V+Q}$:
\begin{eqnarray}
K^2_{m+1}&=&V+Q\nonumber\\
&-&\omega^4\left[\Delta_L(K_m)
+\rmi\gamma_L(K_m)\,\frac{k_m\cot{K_mL}-\rmi K_m}{K_m\cot{K_mL}-\rmi k_m}\right],
\nonumber\\
k_m&=&\sqrt{K^2_m-V}.\label{eqKK}
\end{eqnarray}
Omitting the indexes $m$ and $m+1$ restores the equation equivalent
to $F(\sqrt{K^2-V})=0$.

To extract the resonance part of $S$-matrix, we expand the complex
function $F(k)$ of a real $k$ in vicinity of complex root
$k_{\rm res}=\sqrt{E_{\rm res}}=k_r-\rmi\kappa_r$ as
\begin{equation}
F(k)\simeq C\, (k-k_r+\rmi \kappa_r),\quad
F(k_{\rm res})=0,
\end{equation}
where $C\equiv F^\prime(k_{\rm res})$ is a complex constant.

Since $k=k^*_{\rm res}$ solves conjugate equation $F^*(k)=0$,
and $F^*(k)\simeq C^* (k-k_r-\rmi \kappa_r)$ near the resonance,
we define the phase shift $\delta_0$ related with potential
scattering so that
\begin{equation}
\rme^{2\rmi\delta_0}=\rme^{-2\rmi kL}\,\frac{C^*}{C}.
\end{equation}

The calculated ingredients enable to write down the scattering
matrix and the total phase shift in the form:
\begin{eqnarray}
S&\simeq&\rme^{2\rmi\delta_0}\,\frac{k-k_r-\rmi\kappa_r}{k-k_r+\rmi\kappa_r},\\
\delta&=&\delta_0-\arctan{\frac{\kappa_r}{k-k_r}}.
\end{eqnarray}

Representing these quantities in the conventional form in terms of $E$,
we must expand $F(\sqrt{E})$ in powers of $E$:
\begin{equation}
F(\sqrt{E})\simeq \frac{C}{2k_{\rm res}}\,(E-E_{\rm res}),
\end{equation}
where $C$ is as above. It leads to redefinition
of phase shift~$\delta_0$ because of the relation:
\begin{equation}
\arctan{\frac{\kappa_r}{k-k_r}}-\frac{1}{2}\arctan{\frac{\kappa_r}{k_r}}
=\arctan{\frac{\Gamma_0}{2(E-E_0)}}.
\end{equation}

\begin{figure}
\begin{center}
\includegraphics[width=6cm]{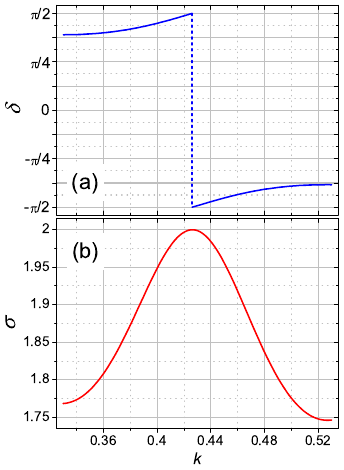}
\end{center}
\vspace*{-7mm}
\caption{\label{fig7}Phase shift $\delta$ (a) and cross section $\sigma$ (b)
as functions of incident momentum $k$. Jump and peak in the graphs
indicate the resonance and appear at
$k_r={\rm Re}\,\sqrt{E_{\rm res}}\simeq0.4275189270$.}
\end{figure}

Thus, we can see that the phase shift $\delta$ experiences a jump
$\delta(k_r-0)-\delta(k_r+0)=\pi$, which reveals a resonance.
Besides, $\delta$ determines the cross section $\sigma$
in accordance with the optic theorem in one dimension~\cite{E65,BM94}:
\begin{equation}
\sigma(k)=2 \sin^2{\delta(k)}.
\end{equation}

Iterating Eq.~(\ref{eqKK}) for $L=5$, $V=0.04$, $Q=0.15$,
and $\omega^4_1=\omega^4_c$, used to test the Feshbach phenomenon
above, we obtain the solution:
\begin{equation}\label{E1}
E_{\rm res}=0.1752355035-\rmi\, 0.07423053097.
\end{equation}

The behavior of $\delta(k)$ and $\sigma(k)$ is depicted in Fig.~\ref{fig7}
and shows that the incident particle with momentum $k=\sqrt{Q}$ can be bound,
if $k$ is within the interval $(k_r-\kappa_r/2;k_r+\kappa_r/2)$ for
$k_r={\rm Re}\,\sqrt{E_{\rm res}}>0$ and $\kappa_r=-{\rm Im}\,\sqrt{E_{\rm res}}>0$.
Note that the asymmetric form of the resonant peak in Fig.~\ref{fig7}(b)
is due to the term $-kL$ in the phase shift $\delta_0$.

The obtained formulas and results describe, in general, the mechanism of
the occurrence of resonance (associated with two-particle
complex -- dimer) without specifying the extra interactions used.
Although the model needs to be refined in accordance with
specific physical conditions, this formalism remains applicable to
various studies.

\subsection{\label{S4E}Analysis of DM Dimer}

Let us now convert dimensionless characteristics into physical units.
The main scale we need is the length scale $r_0$ associated with the
interaction radius. In principle, its refinement requires additional
considerations that are beyond the scope of this study. Anyway, we
apply $r_0$ which is given by Eq.~(\ref{r00}) and is a measure of
the size of the DM halo core. Taking into account the relation to
dimensionless variable in Sec.~\ref{S2}, the nonrelativistic scales
for energy and time are
\begin{equation}
\varepsilon_0=\frac{\hbar^2}{2 m r_0^2},\quad
\tau_0=\frac{\hbar}{2\varepsilon_0},
\end{equation}
where $m$ is the mass of axionlike particle, and $\tau_0$ is determined
using the (minimum) uncertainty principle.

Substituting the typical values of $m=10^{-22}\,\text{eV}/c^2$ and
$r_0=0.138\,\text{kpc}$ from Sec.~\ref{S2}, we arrive at the estimates
\begin{equation}
\varepsilon_0\simeq1.074\times10^{-29}\,\text{eV},\quad
\tau_0\simeq9.715\times10^5\,\text{yrs}.
\end{equation}

Analyzing, the resonance energy
$\varepsilon_0\, {\rm Re}\,E_{\rm res}\simeq1.882\times10^{-30}\,\text{eV}$
turns out to be tens of orders of magnitude lower than the critical temperature
$T^{(d=3)}_c$ of free-boson BEC in three dimensions. It results from substituting
the axionlike particle  concentration
$n^{(d=3)}=\rho_0/m\simeq5.61\times10^{38}\,\text{m}^{-3}$ at typical
parameters in Sec.~\ref{S2} into the known expression written for $d$
dimensions:
\begin{equation}
T^{(d)}_c=\frac{2\pi\hbar^2}{m} \left(\frac{n^{(d)}}{\zeta(d/2)}\right)^{2/d},
\end{equation}
where $\zeta(s)$ is the Riemann zeta-function.

In fact, we get the same result in the one-dimensional case,
defining the concentration as $n^{(d=1)}\sim(n^{(d=3)})^{1/3}$.

Defining the resonance lifetime as $t=2\tau_0/\Gamma_0$ and accounting for
$\Gamma_0\simeq0.148$ as in the test above, we deduce compound system stability
over a period
\begin{equation}
t\simeq1.313\times10^7\,\text{yrs}.
\end{equation}

We associate the resonance with a dimer (two-axion composite), whose
one-dimensional wave function $\chi_{\rm D}(\xi)$ and the binding
energy $E_{\rm bind}$ in dimensionless units are expected to be
[see Eqs.~(\ref{Ebi})-(\ref{kabi})]
\begin{equation}
\chi_{\rm D}(\xi)\propto \exp{\left(-\frac{\xi}{a}\right)},\quad
E_{\rm bind}=-\frac{1}{a^2}.
\end{equation}
Here $a$ is the large scattering length given by Eq.~(\ref{scl1});
$\xi\gg L$ is a dimensionless distance between particles; besides, the
dimensionless binding energy has to be converted into physical units
using the scale $\varepsilon_0$.

These formulas are valid for $a\gg L$, where $L$ is related with
the dimensionless radius of axionlike interaction (see Sec.~\ref{Sec3}).
This indicates the regime of large scattering length $a$, which is
achieved for a coupling $\omega^2$ near its critical value $\omega^2_c$
(see Fig.~\ref{fig6}). Thereby, it confirms the hypothesis about
the need for large~$a$ made in Ref.~\cite{GN22}.

Note that the three-dimensional wave function of a dimer in the spherically
symmetric case behaves as $\exp{(-\mu r)}/r$, where $r$ is the distance
between particles, and is often encountered for a simplified description of
composites in nuclear physics, for example, as deuteron in the ground
state~\cite{BrH06,LL}.
Zoo of diverse (two-particle) states in this field of physics also
enables to discover analogs of molecules, the formation of which is
viewed within the Feshbach resonance concept.
An important guide for confirming the existence of axion dimers
can be dipion molecules.
This follows from the common nature of axions and pions~\cite{PQ77},
that may also lead to dark analogs of pions and their molecules.

In any case, the long-term resonance in our scenario suggests consideration
of DM as multicomponent environment due to the participation of composites.
The presence of composites affects BEC DM properties and stimulates
a detailed study of aspects of the composites formation.

Moreover, the dependence of the pair scattering length on interactions
gives us a theoretical possibility to explain a vanishingly small $a$ observed in
the BEC models with two-particle interaction~\cite{Harko2011,GKN20,GN21}, as
well as its variation in the DM halo of dwarf galaxies.
Indeed, this can be done on the basis of Eq.~(\ref{a22}).
For the sake of correctness, it requires specifying the interaction parameters $V$
and $\omega^2$. Perhaps, the gravity also plays a certain role, which we
explicitly do not take into account here when studying the Feshbach resonance.

\section{\label{S5}Conclusion}

In an attempt to describe the axionlike DM, we have involved a periodic
chiral self-interaction that initially possesses $U(1)$ symmetry, which is broken
in the massive system due to including Newtonian gravity. While the expected two different
phases in a self-gravitating BEC DM are mostly revealed by a jump in the particle density,
the desired multispecies matter, according to the current view, can be formed by
both axionlike particles and their composites created in the processes that are similar
to those in high energy (particle) physics. These phenomena can be helpful
for identifying DM particles.

We develop the effects within the quantum mechanics, finding both the BEC wave function
and the wave functions of both the composite (dimer) and its constituents.
This means that we stay aside the quantized fields with a variable number
of particles, appealing to the quantum mechanical formation and decay of DM dimers.
Since the Gamow's tunneling seems unsuitable even for describing dimer decay, the Feshbach
resonance theory appears to be relevant. Typically, the Feshbach resonance requires
different scattering channels with their own interaction potentials, so one may be
faced with assigning internal degrees of freedom to DM particles, the configuration
of which determines each channel. Ignoring this issue as for now, we only focus on the problem
with additional model potentials. This way, in particular, may be sufficient to describe
the spontaneous creation and decay of compound particles in nuclear physics~\cite{Yam93}.
On the other hand, the detection of Feshbach dimers in atomic BECs in the
laboratory~\cite{Grimm10,BrH06} made it possible to trace the resonance mechanism in detail,
motivating us to apply it to the axionlike DM model and outline its implementation.
An important property of the Feshbach resonance is the increase of the scattering length
to infinity, which coincides with the condition for the creation of composites, disclosed
by the use of effective models~\cite{GN22}.

Thus, in our theoretical study we combine the models of axionlike particles, including
the sine-Gordon equation with its soliton solutions, and the Feshbach resonance concept.
We focused on the quantum-mechanical formation of axion dimers in scattering processes
and discussed in this regard the multicomponent DM consisting of axions and their composites,
which were previously predicted in the qualitative study in Ref.~\cite{GN22}. Besides,
having obtained the crucially important scattering length dependence on the interaction
parameters, we got an idea of how to ensure its very small but different values when
describing the DM halo of various (dwarf) galaxies on the base of
the Gross--Pitaevskii--Poisson equations with pair interaction~\cite{Harko2011,Chavanis2,GN22}.

Using the cosinelike self-interaction derived for QCD axions~\cite{PQ77,Wit80},
we formulated in Sec.~\ref{S2} the gravitating BEC DM model based on
the modified Gross--Pitaevskii--Poisson equations (\ref{feq1})-(\ref{feq2})
at zero temperature. This type of nonlinearity generalizes the polynomial
interactions that were used in the preceding models and are looking as its
truncation~\cite{Harko2011,Chavanis2,GKN20,GN21}.
On the other hand, the gained properties motivated us, first, to look for distinct
phases of the axionlike BEC DM. Indeed, the existence of both rarefied or dilute
(gaseous) and dense (liquidlike) phases immediately results from two independent
solutions to the Gross--Pitaevskii equation even without developing a statistical
description. But, the first-order phase transition in this model, the dynamics of
which has yet to be detailed, would not be independently controlled by
the strengths of pair and three-particle interactions as previously.
This situation is in contrast with that considered in Ref.~\cite{GN21}.
In any case, the phase transition between states must be stimulated
by pressure/compression induced by long-wavelength quantum
fluctuations~\cite{GKN20}. The predominance of one or another phase in
the DM halo of a certain galaxy can be inferred from the characteristics
determined by fitting rotation curves and other observables.
For instance, the DM gaseous (dilute) phase dominates in galaxy M81dwB,
according to Ref.~\cite{GN21}.

Regarding composites of DM, we note that the dense BEC phase is unfavored
for composites because of their probable destruction caused by frequent
collisions, as shown in \cite{GN22}. On the other hand, for their appearance
in a rarefied phase, a large  scattering length is needed and has to be argued.
Although an interaction potential with nonzero asymptotics often leads to
an extremely large scattering length in the Born approximation~\cite{LL},
the scattering length also diverges due to zero-energy resonance, when
the scattered particle goes into a bound state.  Choosing the latter option,
the bound state associated with a composite of at least two particles must be
characterized by an isolated (quasi-discrete) level of positive energy and
a finite lifetime. This is dictated, in particular, by the Feshbach resonance
concept~\cite{Fesh62,Joa75,Yam93}.

Intending to get more analytical results, we turn to the one-dimensional
case. Then, focusing on the problem of a few interacting axions in
the ground state, the three-dimensional Gross--Pitaevskii equation
reduces to the stationary sine-Gordon equation with its antikink solution
as in Sec.~\ref{Sec3}. Comparing Fig.~\ref{chi} and Fig.~\ref{XW}, we see that
the effective potential $W$ in Fig.~\ref{XW} basically inherits the behavior
of the gravitation-modified potentials for the two phases in Fig.~\ref{chi}(b),
while the antikink solution mimics the DM distribution in the DM halo core.
The discrepancy between the particle energies on the right-hand sides
of Eq.~(\ref{sch1b}) and Eq.~(\ref{SG1}) means that the distribution
profile in Fig.~\ref{XW} is formed by axions in the state of zero energy.
Therefore, we are faced with the need to explain the appearance of axions
with zero energy in a one-dimensional trap~$W$, despite the presence of
a domain wall.

To resolve this problem for at least two particles, we use two scattering
channels: closed and open. A closed channel is represented by a bound state
induced by the potential $W$ with asymptotics $W\to1$. An open channel
implies elastic and asymptotically free scattering with a tiny positive energy.
Let a particle perform transit between the channels coupled by an external impact.
Given both the scattering potential~(\ref{Vi}) in an open channel
and the coupling~(\ref{exi}), the two-channel quantum-mechanical problem is
formulated in Sec.~\ref{S4}. Though the square-well potentials are used
therein, another form of them is also allowed. Then, with a certain adjustment of
the parameters of extra interactions, an intermediate level
appears, called the ``dressed'' state, which enables it to overcome
the initial energy gap $Q>0$ between the two channels.
In a sense, such a level appearance is similar to the result of splitting,
within a degeneracy problem under the action of perturbation.

The most significant processes take place in the resonance zone,
which is a finite region of space bounded by a common radius
of  interaction~$L$ (in dimensionless units) for all potentials.
To infer the information about processes far from the resonance
zone, we resort to scattering theory and, thereby, extract data from
the phase shift~$\delta$ of the wave function of outgoing particle
in an open channel, after leaving the resonance zone. Although
the basic scattering characteristics in one and three dimensions
do differ, we relate the scattering length~$a$ with $\delta$
by Eq.~(\ref{adef}) as usual~\cite{BM94}.

The analytical solution (\ref{ss1})-(\ref{ss2>}) of the two-channel
problem is obtained in the first approximation, by taking into account
potential scattering in an open channel with square well~(\ref{Vi}).
This also comprises the characteristics of the dressed state that
occurs when a particle hops between channels with close energies.
Possessing positive energy, the dressed state has a finite
lifetime and the resonance property to decay. If we imagined two
interacting particles, one of which is pinned at the origin, then
the dressed state would be seen as a compound system or an excited
dimer. Besides, one justifies a nonzero decay width at zero energy
due to the dependence of dressed state characteristics (\ref{Rch}) on
the magnitude $V$ of the potential~(\ref{Vi}).

At zero energy, we consider the Feshbach phenomenon to reveal a bound
state by the divergence of the scattering length $a$ at certain (critical)
value of the external influence. Parametrizing the external
interaction~(\ref{exi}) by $\omega^2$, the critical value $\omega^2_c$ is
determined by $\sqrt{Q}$ in Eq.~(\ref{omc}).
That means that the scattering length $a(\omega^2)$ behaves as
$a(\omega^2_c\pm\epsilon^2)\to\pm\infty$ at $\epsilon\to0$, that is
shown in Fig.~\ref{fig6}, and confirms the existence of a bound
(dressed) state. This is valid in both one and three dimensions,
although the divergent and zero scattering lengths have quite opposite
effects on reflectance and transparency in one and three dimensions~\cite{Tar13}.
Note also the similarity of this phenomenon with that for alkali atoms
in the laboratory, when $\omega^2$ is replaced by a magnetic field $B$.
Thus, one can expect the formation of shallow dimers with binding energy
$E_{\rm bind}=-1/a^2$ in dimensionless units at large
scattering length $a\gg L$.

On the other hand, having got the dependence of $a$ on the interaction
parameters, one could reproduce the vanishingly small values of $a$
that take place in the BEC DM models with pair
interaction~\cite{Harko2011,Chavanis2,GN22}.
Although we provide formulas for this, a detailed analysis
is omitted.

To complete the study, the resonance scattering at nonzero energy
is considered, and we find the complex value of resonant energy
$E_{\rm res}=E_0-\rmi \Gamma_0/2$ as a pole of the scattering matrix
for some fixed values of the parameters: $L=5$, $\sqrt{V}=0.2$,
and $\omega^2=\omega^2_c$ at $Q=0.15$.
 One gets $E_0>0$ and $\Gamma_0>0$, in contrast to the typical bound state
with $E_0<0$ and $\Gamma_0=0$. We conclude that an incident particle
with energy $E=Q$ and momentum $k=\sqrt{Q}$ participates in
the resonance in Fig.~\ref{fig7}, because
$E_0-\Gamma_0/2\leq Q\leq E_0+\Gamma_0/2$.

To estimate the dimer lifetime $t=2\tau_0/\Gamma_0$, we use the time scale
$\tau_0=mr^2_0/\hbar$ for nonrelativistic axions with mass
$m\simeq10^{-22 }\,\text{eV}/c^2$. Substituting the scale for the DM halo
core $r_0\simeq0.138\,\text{kpc}$ found in Sec.~\ref{S2}, jointly with
the numerically obtained value $\Gamma_0\simeq0.148$, {\it we get
the encouraging value of lifetime, namely} $t\simeq1.3\times10^7\,\text{yrs}$.
This may be sufficient for dimers participation in forming large DM
structures.  But, the fate of dimers depends on the potentials used, which
can be of gravitational, stochastic, and even (dark) electromagnetic
(due to the desired Primakoff effect~\cite{Prim51}) nature.
A reasonable justification for the extra scattering channel(s)
needs the existence of internal degrees of freedom in DM particles, allowing
them to interact differently.
In this way, two channels in laboratory experiments with atoms result from
energetically different spin configurations in an applied magnetic field.
Meanwhile, in the case of DM particles, the internal degrees of freedom
may be isospin or something else.

Although in DM-dominated galaxies the scattering length takes on different values~\cite{Harko2011},
caused in our model by the effect of additional interactions with situationally
distinct characteristics, its extremely large value, leading to the formation of
dimers, does require fine tuning of certain conditions.
Probably, such tuning may not always occur in all galaxies and appears to be spontaneous.
However, the emergence conditions for the (unitary) regime of infinite scattering length
may be fulfilled during the formation of galaxies along with changes in the
parameters of extra interactions (similar to a magnetic field change in
the laboratory).

In addition, we would like to note the need to take into account long-lived dimers
in the formation of multicomponent DM halos, which ensures BEC stability according
to the results of Ref.~\cite{TBM13}.
Clearly, these issues requires further study, and likewise the production of dimers and
other composites in the environment~\cite{HPR17,BKL18}.

\acknowledgements

Both authors acknowledge support from the National Academy of Sciences of
Ukraine by its priority project ``Properties of the matter at high energies
and in galaxies during the epoch of the reionization of the Universe''
No.~0123U102248.


\end{document}